\documentstyle[12pt]{article}

\def\be{\begin{equation}}
\def\ee{\end{equation}}
\def\beq{\begin{eqnarray}}
\def\eeq{\end{eqnarray}}
\def\s{\sigma} 
\def\G{\Gamma} 
\def\F{_2F_1}
\def\an{analytic}
\def\ac{\an{} continuation}
\def\hf{hypergeometric functions}
\def\ndim{NDIM}
\def\half{\frac{1}{2}}
\def\threehalf{\frac{3}{2}}

\begin{document}
\baselineskip=28pt

\begin{center}
{\large \bf An easy way to solve two-loop vertex integrals}
\end{center}

\begin{center}
Alfredo T. Suzuki\footnote{E-mail:suzuki@ift.unesp.br} and Alexandre G. M. 
Schmidt\footnote{E-mail:schmidt@ift.unesp.br}    
\end{center}

\begin{center}
{\it Instituto de F\'{\i}sica Te\'orica, Universidade Estadual
Paulista, R.Pamplona 145, S\~ao Paulo  SP, CEP 01405-900, Brazil. }
\end{center}

\begin{center}
{\bf Abstract}
\end{center}

Negative dimensional integration is a step further dimensional
regularization ideas. In this approach, based on the principle of
\ac{}, Feynman integrals are polynomial ones and for this reason very
simple to handle, contrary to the usual parametric ones. The result of
the integral worked out in $D<0$ must be analytically continued again
--- of course --- to real physical world, $D>0$, and this
step presents no difficulties. We consider four two-loop three-point
vertex diagrams with arbitrary exponents of propagators and dimension, and two
legs on-shell. As far as we know there is no similar calculation in the
literature, and our original results are neatly expressed in terms of products of gamma
functions.


{\it Running Title: An easy way to solve two-loop vertex integrals}
\vfill\eject

{\bf I. Introduction.}
\vskip.2in

The comprehension and achievements of Quantum Field Theory are mostly
due to the perturbative approach, where the calculation of Feynman
diagrams is an inevitable task. There are several techniques to solve
the associated integrals, wherein the simpler the method to solve them
the better. We can just mention a few ones: the Mellin-Barnes'
representation of massive propagators\cite{boos}, the Gegenbauer
polynomial approach in configuration space\cite{russo} and some
others\cite{kreimer}.    

But none of them use the principle of \ac{} in such an interesting and
beautiful way as the technique\cite{halliday,suzuki} known as Negative
Dimensional Integration Method(\ndim{}). Moreover, \ndim{} gives {\it
simultaneously} several results and shares all the concepts of
dimensional regularization\cite{thooft}. We do not know other approach
which has this amazing property.

Our aim here is to present original results for four scalar integrals
pertaining to three-point vertex diagrams at two-loop level where two of the
external legs are set on-shell. They are
original in the sense that all the exponents of propagators and the 
space-time dimension are left arbitrary, and as far as we know there is no
similar calculation in the literature. 

The outline for our paper is as follows: in section 2 we perform the
Feynman integrals in the negative dimension integration approach, i.e.,
solve them in $D<0$, and then analytically continue the result to real
physical world, $D>0$. In section 3, we consider the special cases of
interest and then in section 4 we present our conclusion.

\vskip.3in
{\bf II. Feynman Graphs with Five and Four Massless Propagators.}
\vskip.2in

To calculate the scalar integral that contributes to the Feynman
diagram of figure 1, we will follow the procedure of \cite{lab}. Let
the gaussian integral be: 

\beq \label{gauss3} G_1(p^2;D) &=& \int d^D\!r\;d^D\!q \exp{\left[
-\alpha q^2-\beta r^2- \gamma(p-r-q)^2-\omega (q+k)^2\right.}\nonumber\\ 
&& \left. -\theta(q+k+r)^2\right], \eeq
which, after some simple algebra we get as a result,
\be \label{gauss3b} G_1(p^2;D) = \left(\frac{\pi^2}{\mu}\right)^{D/2}
\exp{\left(-\frac{1}{\mu} \alpha\beta\gamma p^2\right)} ,\ee
where $\mu=\alpha\beta+\alpha\gamma+\alpha\theta+\beta\gamma
+\beta\theta+ \beta\omega+ \theta\omega+ \gamma\omega$. As in reference
\cite{lab}, we expand (\ref{gauss3b}) in Taylor series, 

\beq G_1(p^2;D) &=& \pi^D \sum_{\{n_j\}=0}^\infty \frac{(-p^2)^{n_1}
(-n_1-\half D)!}{n_1!n_2!n_3!n_4!n_5!n_6!n_7!n_8!n_9!}
\alpha^{n_1+n_2+n_3+n_4}  \nonumber\\
&&\times \beta^{n_1+n_2+n_5+n_6+n_7}\gamma^{n_1+n_3+n_5+n_9}
\theta^{n_4+n_6+n_8} \omega^{n_7+n_8+n_9}, \label{comp1}\eeq
where we used a multinomial expansion for $\mu$ and for this reason, the
constraint 
\[ -n_1-\half D = n_2+n_3+n_4+n_5+n_6+n_7+n_8+n_9 ,\]
must be satisfied. Observe that the indices $\{n_j\}$ take only
positive values, so, $D$ is negative. Consider now the expansion ---
before integration --- in Taylor series of the same gaussian integral,
eq.(\ref{gauss3}), 

\be \label{comp2} G_1(p^2;D) = \sum_{i,j,l,m,n=0}^{\infty}
\frac{(-1)^{i+j+l+m+n}}{i!j!l!m!n!}
\alpha^i\beta^j\gamma^l\omega^m\theta^n {\cal A}(i,j,l,m,n;D),\ee
where
\be \label{B} {\cal A}(i,j,l,m,n;D) \equiv \!\int
d^D\!r\;d^D\!q\;(q^2)^i (r^2)^j (p-r-q)^{2l} (q+k)^{2m} (q+k+r)^{2n},\ee
is the negative dimensional integral. Note that its integrand is formed by
the squares of the internal momenta flowing in the diagram raised to
arbitrary positive integral powers. Were these powers negative, one would
have the usual Feynman integral for the diagram with propagators in
the denominator. Comparing (\ref{comp1}) and (\ref{comp2}) we can solve
for the negative dimensional Feynman integral, ${\cal A}$,

\beq  {\cal A}(i,j,l,m,n;D) &=& \frac{(-\pi)^D (p^2)^{\s}
g}{\G(1+\s)} \sum_{\{n_j\}=0} ^\infty
\frac{\delta_{n_1+n_2+n_3+n_4,i}\delta_{n_1+n_2+n_5+n_6+n_7,j}}
{n_2!n_3!n_4!n_5!n_6!n_7!n_8!n_9!} \nonumber\\  
&&\times \delta_{n_1+n_3+n_5+n_9,l}\delta_{n_4+n_6+n_8,m}
\delta_{n_7+n_8+n_9,n},\eeq   
where $\s=i+j+l+m+n+D$ and 
\[ g = \G(1+i)\G(1+j)\G(1+l)\G(1+m)\G(1+n)\G(1-\s-\half D) .\]  

We conclude that in solving a system of linear algebraic equations we
have the result for the Feynman integral in negative dimensions, which
analytically continued to positive $D$ in a convenient way will reproduce the
standard Feynman integral. The system to be satisfied has five equations and
eight indices (the ``unknowns''),   

\be \left\{ \begin{array}{rcl}
n_2+n_3+n_4&=&i-\s \\
n_2+n_5+n_6+n_7&=&j-\s    \\
n_3+n_5+n_9&=&l-\s \\
n_4+n_6+n_8&=&m \\ 
n_7+n_8+n_9&=&n \end{array} \right. ,\ee 
because one of them reads $n_1=\s$. This means that to solve the system we
have to leave three indices free, or said equivalently, we have to solve
$C_3^8=56$ different systems. Of these, twenty turns out to be unsolvable
systems, or systems where the set solution is empty. Therefore, it remains
thirty-six with non-trivial sets of solutions. However, since these
thirty-six must give the same end result after the summation is performed,
i.e., a thirty-six-fold degeneracy, we need to solve only one of them. 

One such solution is, 
\be
\left \{
\begin{array}{rcl}
n_1&=&\s,\\
n_2&=&-l-m-\half D-n_7,\\
n_3&=&-j-m-n-\half D+n_6+n_7+n_8,\\
n_4&=&m-n_6-n_7-n_8,\\ 
n_5&=&-i-n-\half D-n_6,\\
n_9&=&n-n_7-n_8.
\end{array}
\right.
\ee 

Following the same ideas of \cite{lab} we can sum the three series (it is   
easier to sum first in $n_8$ because the other two, $n_6$, and $n_7$
decouple) and we get,   

\beq {\cal A}(i,j,l,m,n;D) &=& \frac{(-\pi)^D(p^2)^{\s}\G(1+i)\G(1+j)
\G(1+l)}{\G(1+\s)\G(1+i-\s)\G(1+l-\s)\G(1-j-\half D)}
\nonumber\\ 
&&\times \frac{\G(1-\s-\half D)\G(1-j-l-m-D)}
{\G(1-i-n-\half D)\G(1-l-m-\half D)\G(1+\s)} \\
&&\times \G(1-i-j-n-D) \nonumber,\eeq
in negative $D$. Using the property of Pochhammer symbol,
\be  \label{prop} (a|-k) = \frac{(-1)^k}{(1-a|k)} ,\ee
where 
\[ (a|l) \equiv (a)_l = \frac{\G(a+l)}{\G(a)} ,\]
we perform the \ac{}, i.e., bring back the result to real world,
 
\beq \label{resultB} {\cal A}^{\bf AC}(i,j,l,m,n;D) &=& (-\pi)^D(p^2)^{\s}
(-i|\s)(-l|\s)(\s+\half D|-2\s-\half D) \nonumber\\
&& \!\!\times (-j|i+j+n+\half D)(j+l+m+D|-l-m-\half D)\nonumber\\
&&\times (i+j+n+D|-i-j+l+m-n-\half D) .\eeq

This is the result for the Feynman graph of figure 1 with two external
legs on-shell and arbitrary exponents of propagators and dimension.

Consider now the graph of figure 2 and its gaussian integral,

\beq \label{gauss2} G_2(p^2;D) &=& \int d^D\!r\;d^D\!q \exp{\left[
-\alpha q^2-\beta (p-q)^2- \gamma(t+q)^2-\omega r^2\right.}\nonumber\\ 
&& \left. -\theta(t+q-r)^2\right], \eeq
carrying out the momentum integration, 
\be G_2(p^2;D) = \left(\frac{\pi^2}{\phi}\right)^{D/2}
\exp{\left[\frac{-\alpha\beta}{\phi}(\theta+\omega)p^2\right]} ,\ee
where $\phi=\alpha\omega+ \beta\omega+\gamma\omega +\theta\omega+
\alpha\theta+ \beta\theta+ \gamma\theta$. Now, expanding it and
(\ref{gauss2}) in Taylor series, $\phi$ in a multinomial one and
comparing both,  

\beq \label{3pontos} {\cal B}(i,j,l,m,n;D) &\equiv& \int
d^D\!r\;d^D\!q\;(r^2)^i (p-q)^{2j} (t+q)^{2l} (r^2)^m (t+q-r)^{2n}
\nonumber\\ 
&=& (-\pi)^D (p^2)^{\s} g \sum_{\{n_j\}=0} ^\infty
\frac{1}{n_1!n_2!n_3!n_4!n_5!n_6!n_7!n_8!n_9!} ,\eeq 
where $g$ and $\s$ are the same as previously defined. The system to be
satisfied now has 6 equations and 9 indices, i.e., there are in fact 84
systems of linear algebraic equations $(6\times 6)$, and note that we
expect the result to be symmetric (see the diagram) between
$m\leftrightarrow n$, 

\be \left\{ \begin{array}{rcl}
n_1+n_2+n_3+n_7&=&i \\
n_1+n_2+n_4+n_8&=&j \\    
n_5+n_9&=&l \\
n_2+n_3+n_4+n_5+n_6&=&m \\ 
n_1+n_6+n_7+n_8+n_9&=&n \\ 
n_1+n_2&=&\s
\end{array} \right. .\ee 

Again, three indices will remain free, that is, the result provided by \ndim{}
will be written as a triple series. Among the 84 possible solutions 52
are trivial ones, so, it remains to solve thirty-two $6\times 6$
systems. The result, as in \cite{lab}, is degenerate, but now a
thirty-two-fold degeneracy. Solving only one will do, and one such solution
is: 
\be
\left\{
\begin{array}{rcl}
n_1&=&-m-\half D-n_7-n_8-n_9,\\
n_2&=&m+\s+\half D+n_7+n_8+n_9,\\
n_3&=&i-\s-n_7,\\
n_4&=&j-\s-n_8,\\ 
n_5&=&l-n_9,\\ 
n_6&=&m+n+\half D.\end{array} \right.
\ee

Inserting these values in the result (\ref{3pontos}) one verifies that the
series can be written in terms of $\F$ \hf{} of unity argument and for this
reason can be summed\cite{lebedev}, yielding   

\beq {\cal B}(i,j,l,m,n;D) &=& \frac{(-\pi)^D(p^2)^{\s}\G(1+i)\G(1+j)
\G(1+m)}{\G(1+n)\G(1-m-\half D)\G(1-n-\half D)\G(1+i-\s)}
\nonumber\\ 
&\times& \frac{\G(1-\s-\half D)\G(1-m-n-D)}{\G(1+j-\s)\G(1+m+n+ \half
D)\G(1+\s)} ,\eeq    
which, after the convenient analytic continuation to real physical world of
positive $D$, gives

\beq \label{resultA} {\cal B}^{\bf AC}(i,j,l,m,n;D) &=& \pi^D(p^2)^{\s}
(-m|m+n+\half D)(-n|m+n+\half D) (-i|\s) \nonumber\\
&\times& (-j|\s)(m+n+D|-2m-2n-\threehalf D)\nonumber\\
&\times&(\s+\half D|-2\s-\half D).\eeq 

This result is valid for arbitrary exponents of propagators and
space-time dimension, in Euclidean space.

For the next graph, figure 3, we begin with another gaussian integral, 

\beq \label{gauss4} G_3(p^2;D) &=& \int d^D\!r\;d^D\!q \exp{\left[
-\alpha q^2-\beta r^2- \gamma(p-q)^2-\omega (p-r)^2\right.}\nonumber\\ 
&& \left. -\theta(k+r)^2\right]. \eeq
This integral is the simplest one of this work; we get as a result,
\be G_3(p^2;D) = \left(\frac{\pi^2}{\lambda_1\lambda_2}\right)^{D/2}
\exp{\left(-\frac{\alpha\gamma}{\lambda_1} p^2\right)}
\exp{\left(-\frac{\beta\omega}{\lambda_2} p^2\right)}  ,\ee
where $\lambda_1=\alpha+\gamma$ and $\lambda_2=\beta+ \theta+ \omega$.
After carrying out the usual procedure we obtain a system of 7
equations with 7 indices, i.e., there is only one solution that after the
\ac{} is,
\beq \label{resultC}{\cal C}^{\bf AC}(i,j,l,m,n;D) &\equiv& \int
d^D\!r\;d^D\!q\;(q^2)^i (r^2)^j (p-q)^{2l} (p-r)^{2m} (k+r)^{2n}
\nonumber\\ 
&=& \pi^D(p^2)^{\s} (i+l+D|-i-l+m+n-\half D)\nonumber\\
&\times&(-i|-l-\half D)(-l|i+l+\half D)(-j|-m-n-\half D) \nonumber\\ 
&\times&(-m|l+m+\half D)(j+m+n+D|-m-\half D) ,\eeq
where $\s$ is the same as previously defined.

The last scalar integral contributes to a graph, see figure 4, that
shows the so-called overlapping divergences\cite{fairlie} and observe
that it has four propagators. We proceed as above and write down only
the final result,

\begin{equation}  \label{Indim}
{\cal D}(i,j,l,m;D) = \int d^D\!q\;d^D\!r\;(q^2)^i\left[(q-p)^2
\right]^j (r^2)^l\left[(r-q-k)^2\right]^m ,\ee
and the general result
\beq \label{resultD}
{\cal D}^{\bf AC}(i,j,l,m;D) &=& \pi^D(p^2)^{\s'}(-i|\s')(-j|\s')
(-l|-m-\frac{1}{2} D) \nonumber\\
&&\times (-m|\s'-i-j-\frac{1}{2} D)(\s'+\frac{1}{2} D|-2\s'-\frac{1}{2} D) 
\nonumber\\
&&\times (D+l+m|-l-\frac{1}{2} D) ,\eeq
where $\s'=i+j+l+m+D=\s-n$.

\vskip.3in
{\bf III. Special Cases.}
\vskip.2in

The special cases that we deal in this section are related to the Feynman
integrals with unity exponents for the propagators. They correspond exactly
to the cases for which $i=j=l=m=n=-1$. For convenience we use a shorthand
notation $(\{-1\})$ to represent $(i=-1,\,j=-1,\,l=-1,\,m=-1,\,n=-1)$.

The first scalar Feynman integral we carried out, eq.(\ref{B}), yields then

\be {\cal A}^{\bf AC}(\{-1\}) = \pi^D(p^2)^{D-5} \frac{\G^2(D-4)\G(5-D)
\G^2(\half D-2)\G(\half D-1)}{\G^2(D-3)\G(\threehalf D-5)} .\ee 

For the second one the result in the same particular case, reads
\be {\cal B}^{\bf AC}(\{-1\}) = \pi^D(p^2)^{D-5} \frac{\G^2(\half D-1)\G(5-D)
\G^2(D-4)\G(2-\half D)}{\G(D-2)\G(\threehalf D-5)} .\ee 
For the third,
\be {\cal C}^{\bf AC}(\{-1\}) = \pi^D(p^2)^{D-5} \frac{\G^2(\half D-2)\G(2-\half D)
\G^2(\half D-1)\G(3-\half D)}{\G(D-3)\G(D-2)} ,\ee 
and for the last one,
\be 
{\cal D}^{\bf AC}(\{-1\}) = \frac{\pi^D(p^2)^{D-4}\Gamma^2(D-3)\Gamma^2(\frac{1}{2}
D-1) \Gamma(2-\frac{1}{2} D)\Gamma(4-D)}{\Gamma(D-2)\Gamma(\frac{3}{2}
D-4)}.\ee

These are the results in Euclidean space.

\vskip.3in
{\bf IV. Conclusion.}
\vskip.2in

NDIM's amazing feature of transfering the task of solving
parametric integrals to solving systems of linear algebraic equations
allows us to greatly simplify the effort to solve loop integrals. With
two-loop vertex diagrams as examples we worked out rather
cumbersome integrals with ease. The \ac{} of space-time
dimension to {\it negative} values has shown advantages: we 
interpret the \ac{} like in dimensional regularization but solve the
Feynman integrals in a much simpler way because they are polynomial
ones. In this approach there are no cumbersome parametric integrals. We
presented original results for four two-loop three-point scalar
integrals, (see fig.1-4) for arbitrary $D$ and exponents of the
propagators, ({\ref{resultB}), ({\ref{resultA}), ({\ref{resultC}),
({\ref{resultD}). 
\vskip.3in

{\it Acknowledgments}

AGMS gratefully acknowledges CNPq (Conselho Nacional de Desen\-vol\-vi\-men\-to
Cient\'{\i}fico e Tecnol\'ogico, Brazil) for the financial support.

\newpage
\begin{center}
{\large\bf FIGURE CAPTIONS}
\end{center}
\vspace{4cm}

Figure 1: Two-loop three point vertex calculated with NDIM. We
consider that the two lower external legs are on-shell, i.e.,
$k^2=t^2=0$.
\vskip.2in

Figure 2: Another two loop three point graph calculated in the NDIM 
approach.
\vskip.2in

Figure 3: The easiest two loop three point vertex diagram calculated in
the NDIM approach.
\vskip.2in

Figure 4: Two-loop three-point vertex that has overlapping divergences.
\vfill\eject
\newpage

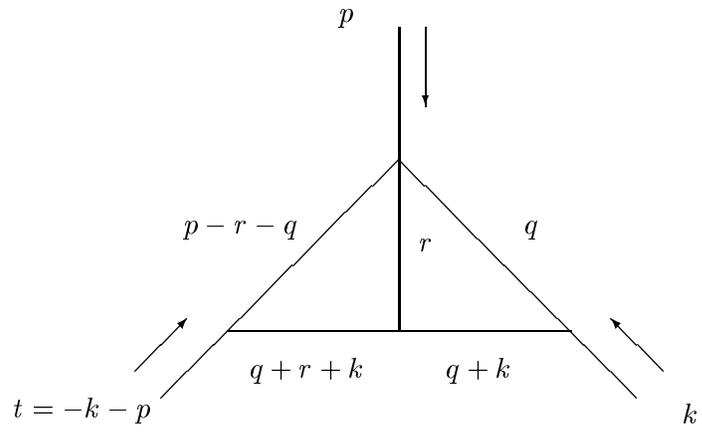
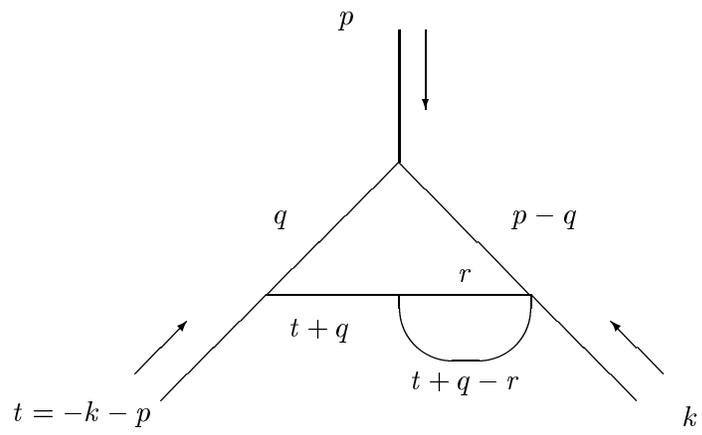
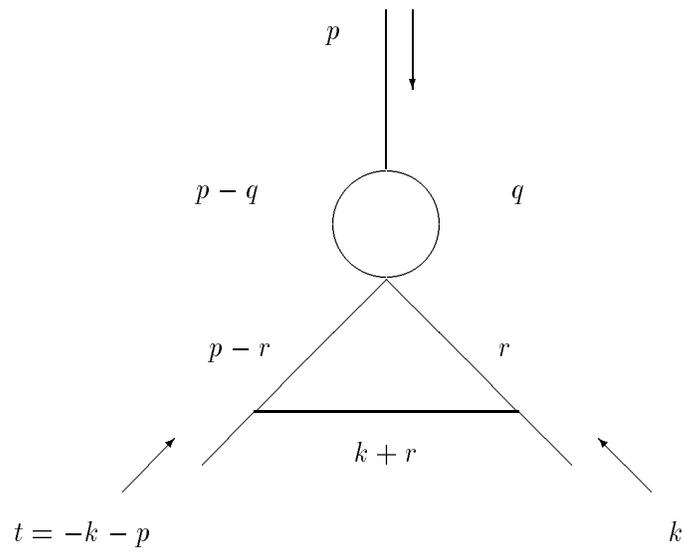
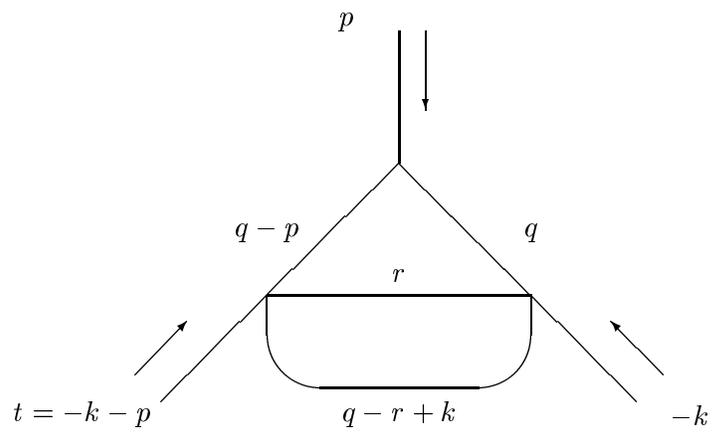
\end{document}